\documentclass[pra,aps,twocolumn,floatfix,showpacs]{revtex4}
\usepackage{graphics}
\begin{document}
\title{Trapped p-wave superfluids: a local density approach}
\author{M. Iskin and C. J. Williams}
\affiliation{Joint Quantum Institute, National Institute of Standards and Technology, 
and University of Maryland, Gaithersburg, Maryland 20899-8423, USA.}
\date{\today}

\begin{abstract}

The local density approximation is used to study the ground state superfluid properties 
of harmonically trapped p-wave Fermi gases as a function of fermion-fermion 
attraction strength. While the density distribution is bimodal on the weakly 
attracting BCS side, it becomes unimodal with increasing attraction 
and saturates towards the BEC side. This non-monotonic evolution is related to the
topological gapless to gapped phase transition, and may be observed via 
radio-frequency spectroscopy since quasi-particle transfer current requires 
a finite threshold only on the BEC side.

\end{abstract}
\maketitle

Recent experiments measuring momentum distribution, collective modes, order parameter, 
quantized vortices, etc. have provided strong evidence for observation of a 
superfluid phase in two-component cold atomic mixtures, interacting with short-range attractive 
s-wave interactions~\cite{djin, rgrimm, jthomas, csalomon, rhulet, mit-vortex}. 
These experiments have also shown evidence that the ground state of 
these s-wave mixtures evolves smoothly from a paired Bardeen-Cooper-Schrieffer (BCS) 
superfluid to a molecular Bose-Einstein condensate (BEC) as the attractive 
interaction varies from weak to strong values, marking the first demonstration 
of theoretically predicted BCS-BEC crossover~\cite{leggett, nsr, jan}.

On the other hand, there has also been substantial experimental progress 
studying p-wave interactions to observe triplet 
superfluidity~\cite{regal-pwave, ticknor-pwave, zhang-pwave, schunck-pwave, gunter-pwave}.
However, controlling the p-wave interactions is proving much 
more difficult due to the narrow nature of p-wave Feshbach 
resonances as well as to the more dramatic two- and also three-body 
losses~\cite{regal-pwave, ticknor-pwave, zhang-pwave, schunck-pwave, gunter-pwave}.
These experiments still motivated considerable theoretical interest 
predicting quantum and topological phase 
transitions~\cite{botelho-pwave, ho-pwave, gurarie-pwave, skyip-pwave, iskin-pwave, stoof-pwave}. 
More recently, p-wave molecules have been produced and their two-body 
properties have been studied~\cite{gaebler-pwave}, opening the possibility 
of studying many-body properties of p-wave superfluids in the near future.

In this manuscript, unlike the previous works on homogenous 
systems~\cite{botelho-pwave, ho-pwave, gurarie-pwave, skyip-pwave, iskin-pwave, stoof-pwave}, 
we study the ground state superfluid properties of harmonically trapped 
p-wave Fermi gases as a function of fermion-fermion 
attraction strength. Our main results are as follows. 
While we find that the density distribution is bimodal on the 
weakly attracting BCS side where the local chemical potentials are positive
everywhere inside the trap, it becomes unimodal with increasing
attraction and saturates towards the BEC side where the local chemical 
potentials become negative. This non-monotonic evolution is related to 
the topological gapless to gapped phase transition occurring in p-wave superfluids, 
and is in sharp contrast with the s-wave case where the superfluid 
phase is always gapped leading to a smooth crossover.
Lastly, we propose that the phase transition found in the p-wave case may be observed 
via radio-frequency spectroscopy since quasi-particle transfer current 
requires a finite threshold only on the BEC side,
which is in sharp contrast with the crossover physics found in the s-wave case 
where a finite threshold is required throughout BCS-BEC evolution.

{\it Local density (LD) approach}:
To obtain these results, we consider a harmonic trapping potential, separate 
the relative motion from the center-of-mass one, and use LD approximation 
to describe the latter. In this approximation, the system is treated as 
locally homogenous at every position inside the trap, and it is valid 
as long as the number of fermions is large which is typically satisfied 
in cold atomic systems~\cite{ohashi-rf}. In order to describe the pairing 
correlations occurring in the relative coordinates, we use the BCS mean 
field (MF) formalism and neglect fluctuations.
The MF description is qualitatively valid throughout BCS-BEC evolution only at the low
temperatures considered here~\cite{leggett, nsr, jan}, and it has been extensively 
applied to cold atomic systems describing qualitatively the experimental observations.

Therefore, we start with the following local MF Hamiltonian (in units of $\hbar = k_B = 1$)
\begin{eqnarray}
\label{eqn:hamiltonian}
H_\ell (r) &=& \sum_{\mathbf{k}, \sigma} \xi_\ell (r, \mathbf{k}) a_{\mathbf{k}, \sigma}^\dagger a_{\mathbf{k}, \sigma}
+ \frac{|\Delta_\ell(r)|^2}{g} \nonumber \\
&-& \sum_\mathbf{k} [\Delta_\ell (r, \mathbf{k}) a_{\mathbf{k}, \uparrow}^\dagger a_{-\mathbf{k}, \downarrow}^\dagger + H.C.],
\end{eqnarray}
where $\ell = 0$ $(\ell = 1)$ corresponds to s-wave (p-wave) systems, 
$a_\mathbf{k, \sigma}^\dagger$ creates a pseudo-spin $\sigma$ fermion with momentum $\mathbf{k}$,
$
\xi_\ell (r, \mathbf{k})= \epsilon (\mathbf{k}) - \mu_\ell (r)
$ 
is the dispersion with
$
\epsilon (\mathbf{k}) = k^2/(2m)
$
and
$
\mu_\ell (r) = \mu_\ell - V (r).
$
While the global chemical potential $\mu_\ell$ fixes the total number of fermions,
the local chemical potential $\mu_\ell (r)$ includes the trapping potential 
$V (r) = m \omega_0^2 r^2/2$.
In Eq.~(\ref{eqn:hamiltonian}), 
$
\Delta_\ell (r, \mathbf{k}) = \Delta_\ell (r) \Gamma_\ell (\mathbf{k})
$ 
is the local MF order parameter where
$
\Delta_\ell (r) = g \sum_\mathbf{k} \Gamma_\ell (\mathbf{k})
\langle a_{-\mathbf{k}, \uparrow} a_{\mathbf{k}, \downarrow} \rangle
$
describes the spatial dependence, such that $g > 0$ is the strengh of the attractive 
fermion-fermion interactions and $\langle ... \rangle$ implies a thermal average.
$\Gamma_\ell (\mathbf{k})$ determines the symmetry of the order parameter given by
$
\Gamma_\ell (\mathbf{k})= W_\ell (k) \sum_m \lambda_{\ell, m} Y_{\ell,m}(\widehat{\mathbf{k}})
$
where
$
W_\ell(k) = k^\ell k_0 /(k^2 + k_0^2)^{(\ell + 1)/2} 
$
describes the momentum dependence~\cite{ho-pwave, iskin-pwave}. Here, $k_0 \sim R_0^{-1}$ sets the 
momentum scale where $R_0$ is the interaction range in real space.
In this manuscript, we assume $\Delta_\ell (r)$ to be real, and consider only 
the $\lambda_{1, 0} = 1$ and $\lambda_{1, \pm 1} = 0$ symmetry. However, we emphasize 
that our discussion applies qualitatively to other p-wave symmetries as well.

The local MF Hamiltonian can be solved by using standard techniques~\cite{ho-pwave, iskin-pwave},
leading to a set of nonlinear equations for $\Delta_\ell (r)$ and $\mu_\ell (r)$.
These equations are
\begin{eqnarray}
\frac{M V}{4\pi a_\ell k_0^{2\ell}} &=& \sum_\mathbf{k} \left[ \frac{W_\ell^2 (k)}{2\epsilon (\mathbf{k})}
- \frac{4\pi \Gamma_\ell^2 (\mathbf{k})}
{2E_\ell (r, \mathbf{k})} \tanh \frac{E_\ell (r, \mathbf{k})}{2T} \right], 
\label{eqn:op} \\
n_\ell (r) &=& \frac{1}{V}\sum_{\mathbf{k}, \sigma} \left[ \frac{1}{2}
- \frac{\xi_\ell (r, \mathbf{k})} {2E_\ell (r, \mathbf{k})} \tanh\frac{E_\ell (r, \mathbf{k})}{2T} \right], 
\label{eqn:number}
\end{eqnarray}
where $a_\ell$ is the experimentally relevant scattering parameter which regularizes $g$,
$
E_\ell (r, \mathbf{k}) = \sqrt{\xi_\ell^2 (r, \mathbf{k}) + \Delta_\ell^2(r, \mathbf{k})}
$
is the local quasi-particle energy, $T$ is the temperature and $V$ is the volume.
In Eq.~(\ref{eqn:number}), $n_\ell (r)$ is the local density of fermions, and the 
total number of fermions $N$ is fixed by $N = \int d\mathbf{r} n_\ell (r)$.
Notice that $a_\ell$ has units of length (volume) in the s-wave (p-wave) case, and 
that our self-consistent solutions also describe the single pseudo-spin
p-wave systems (except for the $\sigma$ summations here and throughout), 
which are presented next.

{\it BCS-BEC evolution in homogenous systems}:
To understand the ground state properties of harmonically trapped p-wave 
superfluids within the LD approach, it is very useful to analyze first 
the homogenous s- and p-wave systems where $V (r) = 0$.
Thus, next we discuss the s-wave case where
$
\Delta_s (\mathbf{k}) = \Delta_s W_0 (k) Y_{0,0} (\widehat{\mathbf{k}})
$
with 
$
Y_{0,0}(\widehat{\mathbf{k}}) = 1/\sqrt{4\pi},
$
and compare these results with the p-wave case where 
$
\Delta_p (\mathbf{k}) = \Delta_p W_1 (k) Y_{1,0}(\widehat{\mathbf{k}})
$
with
$
Y_{1,0}(\widehat{\mathbf{k}}) = \sqrt{3/(4\pi)} \cos(\theta_\mathbf{k}).
$
In the numerical calculations, while we mainly consider $k_0 = 100k_F$ to 
describe realistically the short-ranged atomic interactions, 
some of the $k_0 = 10k_F$ results are also shown for comparison.

\begin{figure} [htb]
\centerline{\scalebox{0.4}{\includegraphics{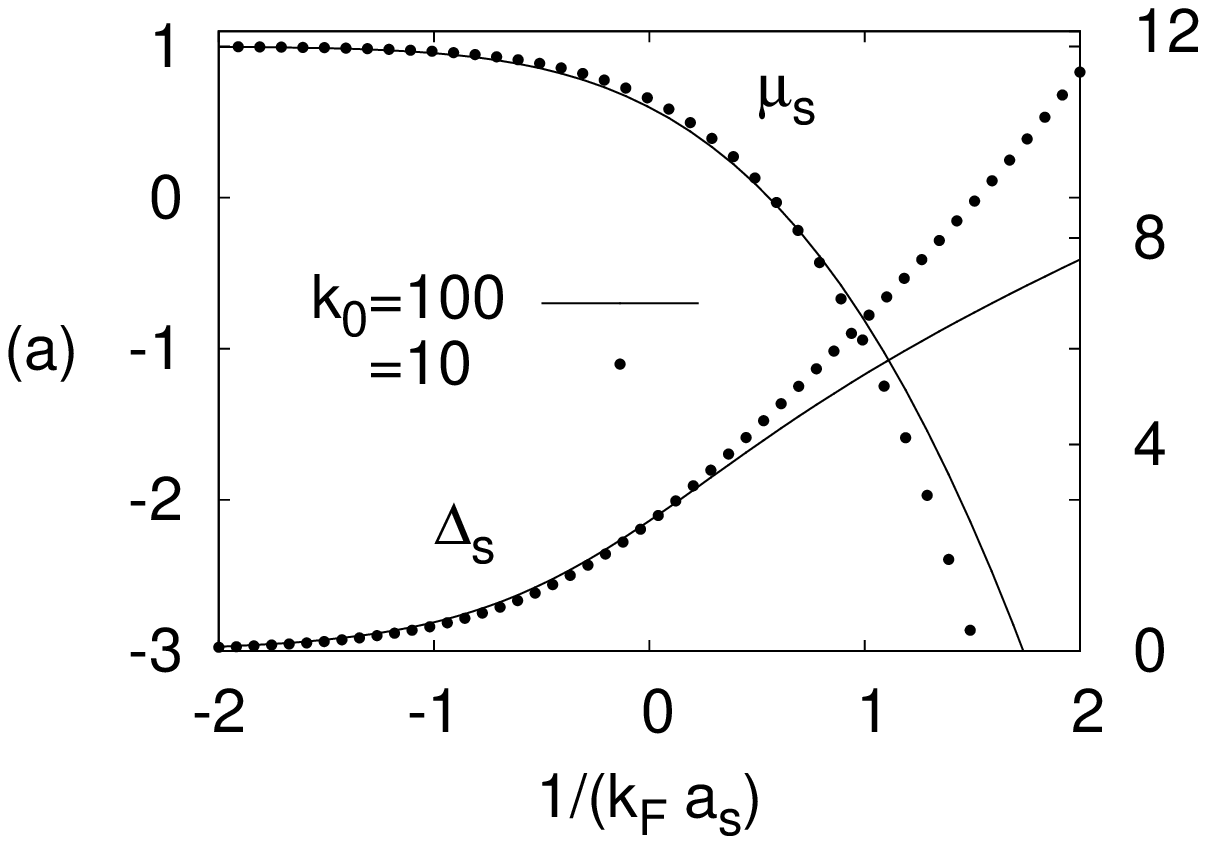}}}
\centerline{\scalebox{0.4}{\includegraphics{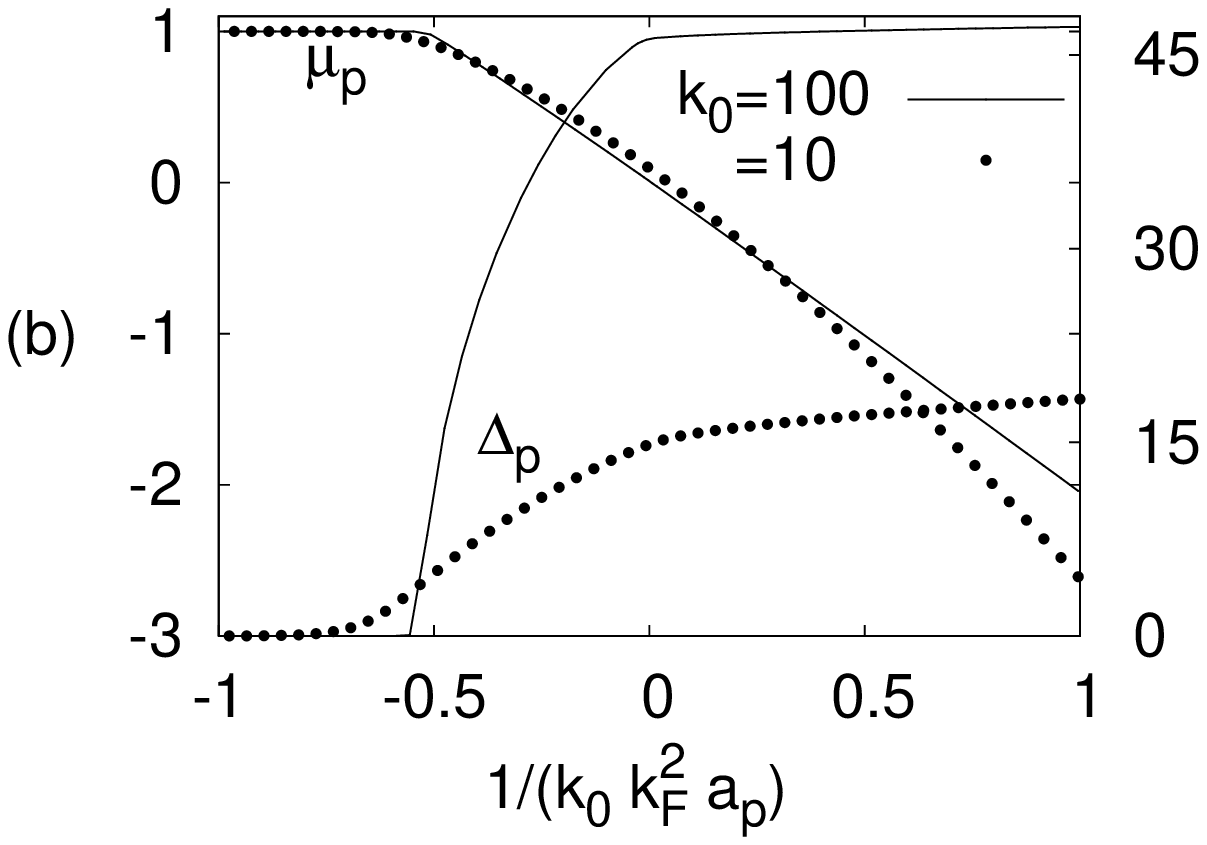}}}
\caption{\label{fig:hom}
We show (in units of $\epsilon_F$)
the chemical potential $\mu_\ell$ (left y-axis) and the amplitude of 
the order parameter $\Delta_\ell$ (right y-axis) for
(a) s-wave systems as a function of $1/(k_F a_s)$, and
(b) p-wave systems as a function of $1/(k_0 k_F^2 a_p)$.
Here, solid (dotted) lines corresponds to $k_0 = 100k_F$ $(k_0 = 10k_F)$.
}
\end{figure}

In Fig.~\ref{fig:hom}(a), we show $\Delta_s$ and $\mu_s$ at zero temperature ($T = 0$) 
for the s-wave case where the BCS-BEC evolution range in $1/(k_F a_s)$ is of 
order $1$. Notice that, $\Delta_s$ grows continuously without saturation with
increasing attraction, while $\mu_s$ decreases continuously from the 
Fermi energy $\epsilon_F = k_F^2/(2m)$ on the BCS side to the half of the binding 
energy $\epsilon_{b,s}/2 = -1/(2m a_s^2)$ on the BEC side~\cite{jan}. 
Here, $k_F$ is the Fermi momentum which fixes the total density $n = \sum_\sigma k_F^3/(6\pi^2)$ 
of fermions. Thus, we conclude that the evolution of $\Delta_s$ and $\mu_s$ as a 
function of $1/(k_F a_s)$ is analytic throughout, and BCS-BEC evolution is a 
smooth crossover~\cite{leggett, nsr, jan}.

In Fig.~\ref{fig:hom}(b), we show $\Delta_p$ and $\mu_p$ at $T = 0$ for the 
p-wave case where the BCS-BEC evolution range in $1/(k_0 k_F^2 a_p)$ is of 
order $1$. Notice that, $\Delta_p$ is exponentially small but still finite 
in the BCS limit when $\mu_p \approx \epsilon_F$ and it grows rapidly with increasing 
attraction but almost saturates for large $1/(k_0 k_F^2 a_p)$, while $\mu_p$ decreases 
continuously from $\epsilon_F$ on the BCS side to $\epsilon_{b,p}/2 = -1/(m k_0 a_p)$ 
on the BEC side. However, both $\Delta_p$ and $\mu_p$ are non-analytic exactly 
when $\mu_p = 0$ at $1/(k_F^3 a_p) \approx 0.45$,
which occurs on the BEC side of unitarity. We note that the 
non-analyticity of $\mu_p$ is barely seen in Fig.~\ref{fig:hom}(b), and it
is more expilicit in derivatives of $\mu_p$.
Thus, in the p-wave case, BCS-BEC evolution is not a crossover, but a quantum 
phase transition occurs~\cite{botelho-pwave, gurarie-pwave, iskin-pwave}.

This phase transition can be understood as follows. The quasi-particle 
excitation spectrum $E_\ell (\mathbf{k})$ is gapless when the conditions 
$\Delta_\ell (\mathbf{k}) = 0$ and $\xi_\ell (\mathbf{k}) = 0$ are both satisfied
for some $\mathbf{k}$-space regions. While the second condition is satisfied for both 
s- and p-wave symmetries on the BCS side where $\mu_\ell > 0$, the first condition is only 
satisfied by the p-wave order parameter. Therefore,  unlike the s-wave case, 
$E_p (\mathbf{k})$ is gapless on the BCS side ($\mu_p > 0$) but it is gapped on 
the BEC side ($\mu_p < 0$), leading to the phase transition discussed
above~\cite{leggett, volovik-book, botelho-pwave, iskin-pwave}.

Having discussed the ground state of homogenous systems,
next we analyze the trapped case.

{\it BCS-BEC evolution in trapped systems}:
For this purpose, similar to the analysis of homogenous systems, 
first we discuss the s-wave case where 
$
\Delta_s (r, \mathbf{k}) = \Delta_s (r) W_0 (k) Y_{0,0} (\widehat{\mathbf{k}}),
$
and compare these results with the p-wave case where 
$
\Delta_p (r, \mathbf{k}) = \Delta_p (r) W_1 (k) Y_{1,0}(\widehat{\mathbf{k}}).
$
In the numerical calculations, we again choose $k_0 = 100k_F$ where 
$
k_F = m \omega_0 r_F
$
is the global Fermi momentum.
Here, $r_F$ is the Thomas-Fermi radius determined by
$
V(r_F) = \epsilon_F = k_F^2/(2m),
$
and fixes the total number of fermions to $N = \sum_\sigma k_F^3 r_F^3/48$.

\begin{figure} [htb]
\centerline{\scalebox{0.37}{\includegraphics{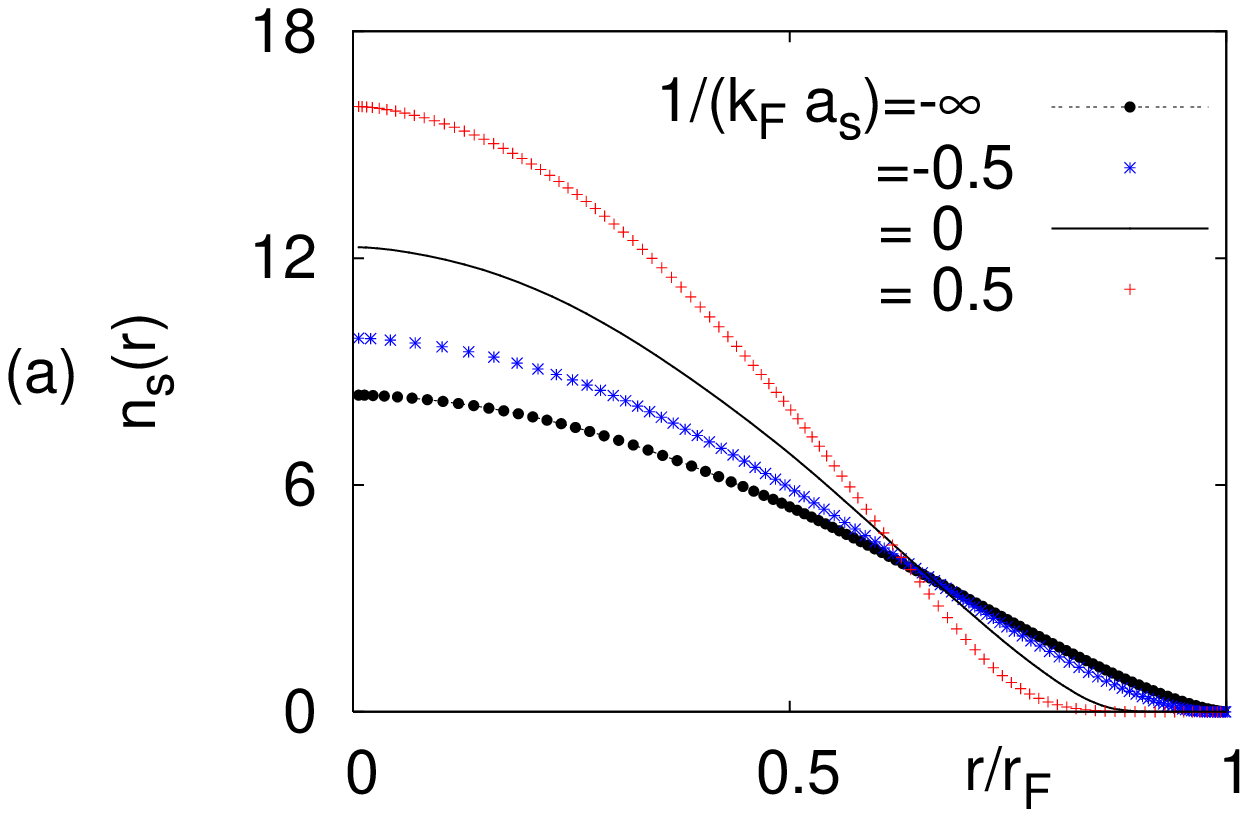}}}
\centerline{\scalebox{0.38}{\includegraphics{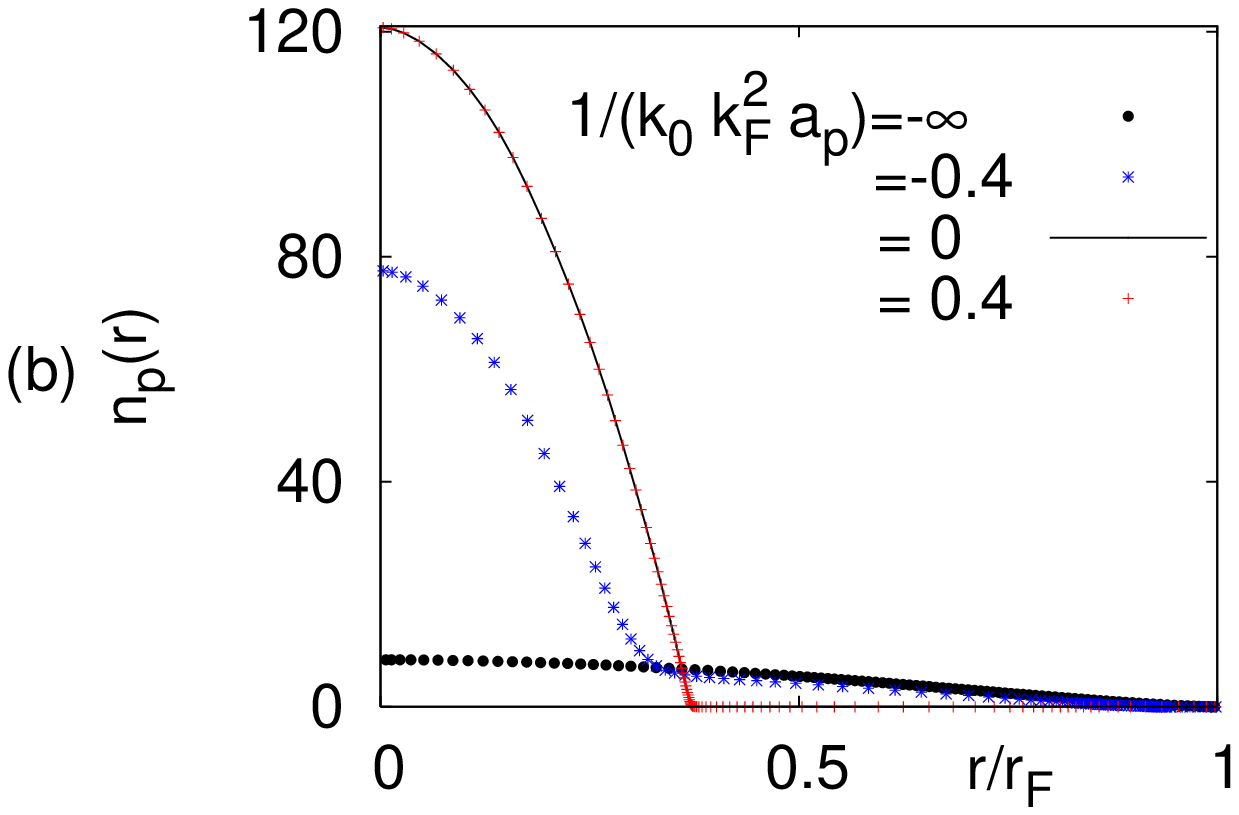}}}
\caption{\label{fig:trap} (Color online)
We show [in units of $k_F^3/(2\pi)^3$] the density distribution $n_\ell (r)$ for
(a) s-, and
(b) p-wave systems
as a function of the trap radius $r$ (in units of $r_F$).
Here, we set $k_0 = 100k_F$.
}
\end{figure}

Within the LD approximation, the density distribution of trapped 
non-interacting ($g = 0$ or $a_\ell \to 0^-$) gas is
$
n_\ell (r) = \sum_\sigma k_F^3(r)/(6\pi^2),
$
where $k_F (r)$ is the local Fermi momentum determined by
$
\mu_\ell = k_F^2(r)/(2m) + V(r)
$
with $\mu_\ell = \epsilon_F$. Therefore, both $k_F (r)$ and $n_\ell (r)$ are 
highest at the center of the trap as can be also seen in Figs.~\ref{fig:trap}(a) 
and~\ref{fig:trap}(b) when $1/(k_F a_s) = -\infty$ 
and $1/(k_0 k_F^2 a_p) = -\infty$, respectively.

In the presence of weak attraction, while $\mu_\ell$ deviates from $\epsilon_F$, 
the density distribution is still well described by the non-interacting 
expression. For fixed $N$, $n_\ell (r)$ is expected to squeeze 
and increase towards the center of the trap since $\mu_\ell$ decreases with 
increasing attraction. The squeezing effect of the weak attractive 
interactions can be seen in Fig.~\ref{fig:trap}(a) for the s-wave and in 
Fig.~\ref{fig:trap}(b) for the p-wave systems when $1/(k_F a_s) = -0.5$ and 
$1/(k_0 k_F^2 a_p) = -0.4$, respectively.

However, weakly attracting p-wave superfluids have bimodal density 
distribution as shown in Fig.~\ref{fig:trap}(b), which is in sharp 
contrast with the unimodal s-wave distribution of Fig.~\ref{fig:trap}(a).
This difference can be understood from the homogenous results shown 
in Fig.~\ref{fig:hom}(a) and~\ref{fig:hom}(b) as follows. 
First, since $k_F (r)$ decreases away from the center of the trap, 
the local s- and p-wave scattering parameters 
$1/[k_F(r) a_s]$ and $1/[k_0 k_F^2(r) a_p]$, respectively, increase
as a function of $r$ if $a_s$ and $a_p$ are fixed.
Second, notice in Fig.~\ref{fig:hom}(b) that $\Delta_p$ increases 
rapidly from exponentially small to larger values as a function
of $1/(k_0 k_F^2 a_p)$, unlike $\Delta_s$ which increases smoothly 
as shown in Fig.~\ref{fig:trap}(a). These two observations combined 
shows that the almost non-interacting $n_p (r)$ distribution towards 
the tail is due to finite but exponentially small $\Delta_p (r)$.

With increasing attraction towards unitarity, while the unimodal $n_s (r)$ 
distribution smoothly squeezes further as shown in Fig.~\ref{fig:trap}(a) 
for $1/(k_F a_s) = 0$ and $1/(k_F a_s) = 0.5$, the bimodal $n_p (r)$ 
distribution becomes unimodal and saturates as shown in Fig.~\ref{fig:trap}(b) 
for $1/(k_0 k_F^2 a_p) = 0$ and $1/(k_0 k_F^2 a_p) = 0.4$. 
This difference can be understood at best on the BEC side where 
strongly attracting fermion pairs form weakly repulsive local molecules which 
can be well described by the Bogoliubov theory~\cite{jan, iskin-pwave}. 
On this side, the size of the s-wave molecules decreases to arbitrarilly 
small values as $\xi_{B, s} \sim a_s > 0$ when $k_0 a_s \gg 1$,
leading to arbitrarilly weak molecule-molecule repulsion 
$U_{BB, s} = 2\pi a_{BB, s}/m$ where $a_{BB, s} = 2a_s$ within the Born approximation~\cite{jan}.
However, the size of the p-wave molecules saturates to small but finite 
values as $\xi_{B, p} \sim 1/k_0$ when $k_0^3 a_p \gg 1$, leading also to a 
weak but finite molecular repulsion $U_{BB, p} = 2\pi a_{BB, p}/m$ 
where $a_{BB, p} = 9/k_0$~\cite{iskin-pwave}. 
Since our LD approximation recovers the Thomas-Fermi approximation 
for the resultant molecules in the BEC limit, $n_p (r)$ saturates rapidly 
for $1/(k_0 k_F^2 a_p) > 0$ due to the presence of weak but finite 
molecular repulsion.

\begin{figure} [htb]
\centerline{\scalebox{0.25}{\includegraphics{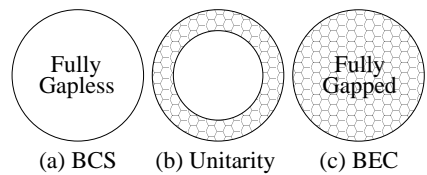}}}
\caption{\label{fig:topology}
Schematic diagrams showing
(a) a fully gapless superfluid on the BCS side 
when $\mu_p (0 \le r \lesssim r_F) > 0$,
(b) a partially gapped superfluid around unitarity 
when $\mu_p (0 \le r \le r_*) > 0$ but $\mu_p (r > r_*) < 0$, and
(c) a fully gapped superfluid on the BEC side 
when $\mu_p (r \ge 0) < 0$.
}
\end{figure}

This non-monotonic evolution of $n_p (r)$ is also related to the 
topological phase transition discussed above for the homogenous systems.
In the trapped case, the local quasi-particle excitation spectrum 
$E_\ell (r, \mathbf{k})$ at position $r$ is gapless when the conditions 
$\Delta_\ell (r, \mathbf{k}) = 0$ and $\xi_\ell (r, \mathbf{k}) = 0$ 
are both satisfied for some $\mathbf{k}$-space regions. 
While these conditions are both satisfied everywhere inside the trap leading to a fully 
gapless superfluid on the BCS side, they are only satisfied around the center of the 
trap close to unitarity leading to a partially gapped superfluid.
Further increasing the attraction towards the BEC limit, the second condition
is not satisfied, and the entire superfluid becomes fully gapped. 
These phases are schematically shown in Fig.~\ref{fig:topology}(a),~\ref{fig:topology}(b) 
and~\ref{fig:topology}(c), respectively, and next we discuss their 
experimental detection.

{\it Radio-frequency (RF) spectroscopy}:
The gapless to gapped phase transition discussed above may be observed for the first
time in cold atomic systems via for instance RF spectroscopy,
where atoms are transferred from one hyperfine state to another generating a
quasi-particle current~\cite{torma-rf, ohashi-rf, chin-rf, shin-rf}. 
This is analogous to electrons tunneling from a superconducting to normal metal, 
and it has been used in atomic systems to observe pairing correlations in 
unpolarized~\cite{chin-rf} as well as polarized~\cite{shin-rf} mixtures.

\begin{figure} [htb]
\centerline{\scalebox{0.37}{\includegraphics{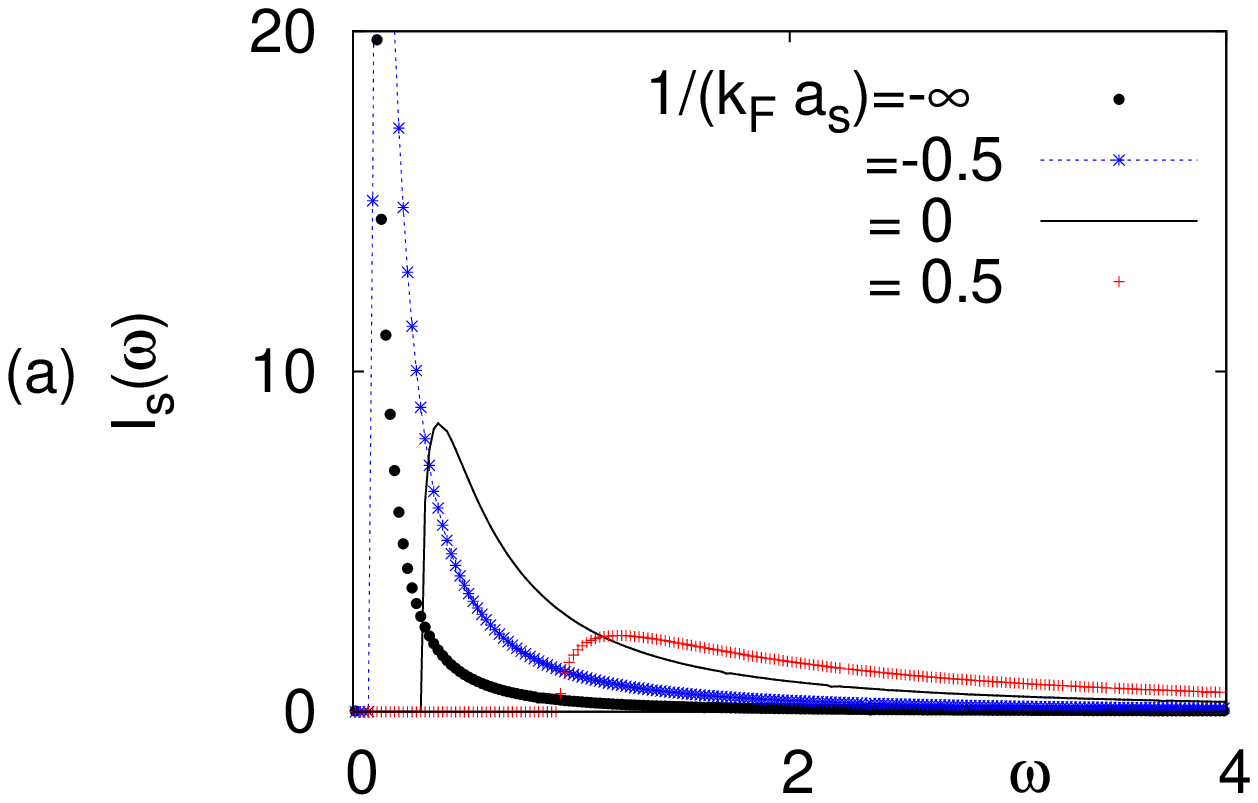}}}
\centerline{\scalebox{0.38}{\includegraphics{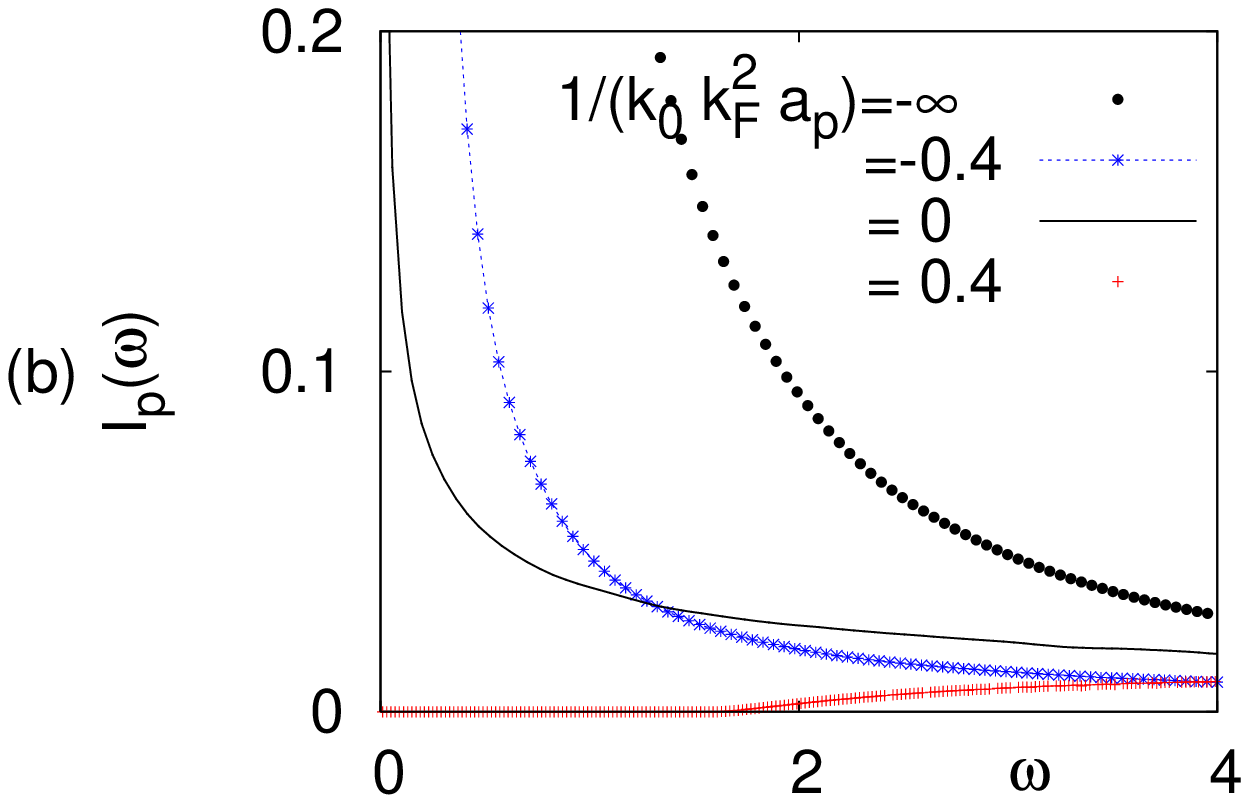}}}
\caption{\label{fig:current} (Color online)
We show (in units of $\rho_F t_F^2 /2$) the quasi-particle transfer
current $I_\ell (\omega)$ for homogenous
(a) s-, and
(b) p-wave systems
as a function of the effective detuning $\omega$ (in units of $\epsilon_F$).
Here, we set $k_0 = 100k_F$.
}
\end{figure}

The local quasi-particle transfer current, within the LD approximation, 
is given by~\cite{mahan-rf, torma-rf, ohashi-rf}
\begin{equation}
\label{eqn:current}
I_\ell (r, \omega) = t_F^2 \sum_{\mathbf{k}} 
A_\ell[\mathbf{k}, \xi_\ell(r, \mathbf{k}) - \omega] F[\xi_\ell(r, \mathbf{k}) - \omega],
\end{equation}
where $t_F$ is the transfer amplitude, $A_\ell(\mathbf{k}, x)$ is the spectral function 
corresponding to the superfluid state, and $F(x) = 1/[\exp(x/T) + 1]$ is the Fermi function. 
Here, $\omega = \omega_L - \omega_H$ is the effective detuning where $\omega_L$ 
and $\omega_H$ are RF laser frequency and hyperfine splitting, respectively.
We evaluate Eq.~(\ref{eqn:current}) with the standard BCS spectral functions 
$
A_\ell (\mathbf{k}, \epsilon) = 2\pi \{ u_\ell^2(r, \mathbf{k}) \delta[\epsilon - E_\ell (r, \mathbf{k})] 
+ v_\ell^2(r, \mathbf{k}) \delta[\epsilon + E_\ell (r, \mathbf{k})]\},
$
where $u_\ell^2(r, \mathbf{k}) = 0.5[1 + \xi_\ell (r, \mathbf{k})/E_\ell (r, \mathbf{k})]$ and
$v_\ell^2(r, \mathbf{k}) = 0.5[1 - \xi_\ell (r, \mathbf{k})/E_\ell (r, \mathbf{k})]$ 
are coherence factors, and $\delta(x)$ is the delta function.

At $T = 0$, the s-wave current $I_s (r, \omega)$ can be evaluated analytically leading to
$
I_s (r, \omega) = \pi \rho_F t_F^2 [\Delta_s (r)/(4\pi \omega)]^2 
\sqrt{C (r, \omega)} \theta (\omega) \theta [C (r, \omega)],
$
where $\rho_F = m V k_F/(2\pi^2)$ is the density of states, 
$C (r, \omega) = [\omega^2 - \Delta^2(r)/(4\pi)]/(2\omega) + \mu_s(r)$,
and $\theta(x)$ is the theta function. Therefore, $I_s (r, \omega)$ flows when the threshold
$\omega_{th, s} (r) \ge - \mu_s(r) + \sqrt{\mu_s^2(r) + \Delta_s^2(r)/(4\pi)} \ge 0$ is reached, which 
reduces to $\omega_{th, s} (r) \ge \Delta_s^2(r)/[8\pi \mu_s(r)]$ in the BCS 
and $\omega_{th, s} (r) \ge 2|\mu_s(r)|$ in the BEC limit. 
Therefore, $w_{th, s} (r) \ne 0$ everywhere inside the trap throughout 
BCS-BEC evolution~\cite{torma-rf, ohashi-rf, chin-rf, shin-rf}.
These finite detuning thresholds can be also seen in Fig.~\ref{fig:current}(a), 
where we show $I_s (\omega)$ for homogenous systems.

The p-wave current $I_p (r, \omega)$ is difficult to evaluate analytically.
However, unlike the s-wave case, we expect that $w_{th, p} (r) \ne 0$ everywhere 
inside the trap only on the BEC side beyond unitarity where $E_p (r, \mathbf{k})$ 
is gapped, and also that $w_{th, p} (r) = 0$ everywhere inside the trap on the BCS side
where $E_p (r, \mathbf{k})$ is gapless. 
The absence (presence) of finite thresholds in gapless (gapped) superfluids 
can be seen in our homogenous results shown in Fig.~\ref{fig:current}(b).
Notice that, while $\omega_{th, p} = 0$ for $1/(k_0 k_F^2 a_p) = -\infty$, $-0.4$ 
and $0$, a finite threshold is required for $1/(k_0 k_F^2 a_p) = 0.4$.
Based on these results, we hope that spatially resolved RF spectroscopy 
measurements (similar to~\cite{shin-rf}) may be used to identify all three 
phases proposed in Fig.~\ref{fig:topology}.

{\it Conclusions}:
To summarize, we showed that while the density distribution of p-wave systems 
is bimodal on the weakly attracting BCS side, it saturates and becomes unimodal 
with increasing attraction towards the BEC side. 
We discussed that this non-monotonic evolution is related to 
the topological gapless to gapped phase transition occurring in p-wave 
superfluids, and is in sharp contrast with the s-wave case where 
the superfluid phase is always gapped leading to a smooth crossover.
Lastly, we proposed that this phase transition may be observed via RF 
spectroscopy since quasi-particle transfer current requires a finite threshold
only on the BEC side, which is in sharp contrast with the s-wave case where 
a finite threshold is required throughout BCS-BEC evolution.

\end{document}